\documentclass[twoside]{ilcws08}         
\usepackage[latin1]{inputenc}
\usepackage[dvips]{graphicx,epsfig,color}
\usepackage{wrapfig,rotating}
\usepackage{amssymb,amsmath,array}
\usepackage{wasysym}  

\newcommand{\ALR}    {\ensuremath{ A_{\mathrm{LR}}} }
\newcommand{\ppl}    {\ensuremath{{\mathcal P}_{e^+}} }
\newcommand{\pmi}    {\ensuremath{{\mathcal P}_{e^-}} }

\newcommand{\peff}   {\ensuremath{{\mathcal P}_{\rm{eff}}} }

\begin{document}
\title{
Beam Polarization at the ILC: the Physics Impact and the
Accelerator Solutions} 
\author{
  B.~Aurand$^{1}$,         
  I.~Bailey$^{2}$,         
  C.~Bartels$^{3}$,        
  A.~Brachmann$^{4}$,      
  J.~Clarke$^{5}$,         
  A.~Hartin$^{3,6}$, \\      
  J.~Hauptman$^{7}$,       
  C.~Helebrant$^{3}$,      
  S.~Hesselbach$^{8}$,     
  D.~K\"afer$^{3}$,        
  J.~List$^{3}$,           
  W.~Lorenzon$^{9}$,   \\  
  I.~Marchesini$^{3}$,     
  K.~M\"onig$^{3}$,        
  K.C.~Moffeit$^{4}$,      
  G.~Moortgat-Pick$^{8}$,  
  S.~Riemann$^{3}$,\footnote{Corresponding author: sabine.riemann@desy.de}  \\ 
  A.~Sch\"alicke$^{3}$,    
  P.~Sch\"uler$^{3}$,       
  P.~Starovoitov$^{10}$,    
  A.~Ushakov$^{3}$,        
  U.~Velte$^{3}$,          
  J.~Wittschen$^{1}$,      
  M.~Woods$^{4}$           
  \vspace{4mm} \\
  1- Phys. Inst., University of Bonn, Germany;\quad         
  2- Cockcroft Inst., University of Liverpool, UK; \\
  3- DESY, Hamburg and Zeuthen, Germany;\quad 
  4- SLAC, Stanford, USA;\quad \\
  5- Daresbury Laboratory, UK;\quad 
  6- JAI, Oxford, UK;\quad \\  
  7- Iowa State University, USA; 
  8- IPPP, University of Durham, UK;\quad \\ 
  9- University of Michigan, USA; 
  10- NCPHEP, Minsk, Belarus;\quad 
}

\maketitle

\vspace*{2mm}
\begin{abstract}
In this contribution accelerator solutions for polarized beams and their impact on physics measurements are discussed.  Focus are physics requirements for precision polarimetry near the interaction point and their realization with polarized sources.
Based on the ILC baseline programme as described in the 
Reference Design Report (RDR), recent developments are discussed and evaluated taking into account physics runs at beam energies 
between 100\,GeV and 250\,GeV, as well as calibration runs on the Z-pole and options as the 1\,TeV upgrade and GigaZ.
\end{abstract}

\section{Introduction}
With the start of LHC operation the physics frontier will be opened to very high energies. 
The full understanding of the results requires completion  by precision measurements  with 
a lepton collider. 
Due to parity violation in weak interactions  
polarized beams are essential to unravel  new phenomena and  
will play a crucial role in that programme. 

\noindent
The beam  polarization  and its importance for physics at an e$^+$e$^-$ collider has been discussed in details over years. A comprehensive overview on the physics prospectives with polarized beams can be found in reference~\cite{ref:power}. 
The focus of this contribution is beam polarization and 
its precise measurement. As precision luminosity and energy measurement, the performance of polarimeters has to be implicated in the the machine design from the beginning to achieve the physics goals for a future lepton collider as the ILC. 
\noindent
The baseline configuration for the ILC is described in the Reference Design Report (RDR)~\cite{bib:RDRacc}. The  electron beam will be highly polarized: 
based on the experience of SLAC, at least 80\% polarization are expected. 
The polarization near the collision will be measured with Compton polarimeters positioned upstream and downstream the interaction point (IP). The complementarity of these two polarimeters
as well as the flexibility of the suggested solutions for various ILC energies
 will be discussed in section~\ref{sec:Pol-IP}.

\noindent
In compliance with the RDR~\cite{bib:RDRacc}, the positron source  configuration of the ILC is based on the helical undulator and  provides a polarized positron beam. 
The degree of positron beam polarization is smaller than planned for the upgrade, about 30\% -- 45\%, but sufficient to be used for physics measurements.

\noindent
In order to preserve the polarization, the spin vectors have to be aligned  parallel to the rotation axis of the damping ring. Hence, in the electron and positron line spin rotator systems \cite{ref:spin-rot}  rotate the spin vector from the longitudinal to the vertical direction before the Damping Ring (DR) and back after.
However,  there is one decisive feature: the helicity of the electrons can be chosen on a train-by-train basis by switching the laser polarization. 
 The helicity of positrons is defined by the orientation of the helix winding in the undulator and additional instrumentation is needed to provide fast helicity reversal also for positrons. The fact that this instrumentation is not yet included in the baseline design initiated discussions  whether fast helicity reversal would be really needed taking into account also the costs.
The consequences of fast or slow helicity reversal 
will be pointed out in in section~\ref{sec:phys}. 

\noindent
Recently the high precision polarization measurement near the IP has been compromised by proposals to combine machine instrumentation and protection equipment with the polarimeter chicane. The impact of such design on polarimetry has been discussed at this and previous workshops, details can be found in references~\cite{ref:jenny-pol-lcws08,ref:EP08}. Here in this contribution,  the basic polarimeter design required for appropriate and reliable ILC  polarization measurements will be emphasized.

%
\section{The Precision  Requirements}{\label{sec:phys}}
The strong potential of the ILC is the precision: Standard Model processes as fermion pair production, W$^+$W$^-$ and also the Higgs production, will be obtained with high statistics and allow uncertainties at the per-mille level. This demands measurements of energy and luminosity with high stability and with precision at a level ${\cal O}(10^{-4})$ as mentioned in the physics part of the RDR~\cite{ref:RDRphys}.
The challenge of  polarization becomes clear 
by comparing uncertainties in  statistics, luminosity and energy with the precision that can be achieved for polarization measurements:
\begin{equation}
\frac{\Delta \sigma}{\sigma} \propto \frac{1}{\sqrt{N}}\oplus \frac{\Delta {\cal
{L}}}{{\cal{L}}}\oplus \frac{\Delta E}{E} \oplus \frac{\Delta P}{P}\,.
\end{equation}
 Studies also presented at this workshop~\cite{ref:pol08} show that a precision of $\Delta P/P\approx 0.25\%$ is possible (see also section~\ref{sec:Pol-IP}).

\noindent 
There are three methods to measure polarization of colliding beams:
upstream and downstream of the IP, as well as using annihilation events. 

\subsection{Basic Remarks} \label{sec:basics}
In general, the cross section for s-channel processes can be 
written as 
\begin{eqnarray}
\sigma_{\pm \pm} & = & \frac{1}{4} \sigma_0 \left[ 1-|P_{e^+}||P_{e^-}| +\ALR 
\left( \mp |P_{e^+}| \pm |P_{e^-}| \right) \right] \label{eq:nonSM} \\
\sigma_{\mp \pm} & = & \frac{1}{4} \sigma_0 \left[ 1 +|P_{e^+}||P_{e^-}| +\ALR
\left(\pm |P_{e^+}| \pm |P_{e^-}| \right) \right] \label{eq:SM} 
\end{eqnarray}
where `$+~-$', `$-~-$', `$-~+$' and `$+~+$' are the different combinations of helicities in the initial state, $\ALR$ the Left-Right asymmetry and $\sigma_0$ the unpolarized cross section. In the Standard Model, the cross sections $\sigma_{\pm \pm}$ are zero for 100\% polarized beams.
The  asymmetry, $\ALR$, can be determined from the
 measured left-right asymmetric cross sections with
\begin{equation}
\ALR= \frac{\sigma_{- +}- \sigma_{+ -}}{\sigma_{- +}+ \sigma_{+ -}} \cdot
       \frac{1-P_{e^+}P_{e^-}} {P_{e^+}-P_{e^-}}=\frac{\ALR^{meas}}{P_{eff}}\label{eq:alr1}\,,
\end{equation} 
and the unpolarized cross section is
\begin{equation}
\sigma_0 = \frac{1}{2} \cdot \frac{\sigma_{- +} +\sigma_{+ -}}{1-P_{e^+}P_{e^-}}
\,.
\end{equation} 
In case of electron polarization only,  these observables simplify to 
\begin{equation} 
\sigma_{\mp} = \sigma_0\left[1 \pm \pmi\,\ALR \right]\,, ~~~~~~~
\ALR= \frac{\sigma_- - \sigma_+ }{\sigma_- + \sigma_+ } \cdot
       \frac{1} {P_{e^-}} = \frac{\ALR^{meas}}{P_{e}}
\end{equation}
The event numbers, $N_{\mp \pm}$ and $N_{\pm \pm}$ are measured if the luminosities ${\cal{L}}_{\mp \pm}$, ${\cal{L}}_{\pm \pm}$,  are delivered to the different initial states with the polarizations $(\mp |P_{e^-}|, \pm |P_{e^+}|)$ and  $(\pm |P_{e^-}|, \pm |P_{e^+}|)$.
If the luminosity is equally  distributed to the '$- ~ +$' and '$+ ~ -$' initial states, ${\cal{L}}_{\mp \pm} = {\cal{L}}_{ \pm \mp}$,  the Left-Right asymmetry is
\begin{equation}
 \ALR=\frac{N_{- +} - N_{+ -}}{N_{- +} + N_{+ -}}\cdot \frac{1}{P_{eff}}\label{eq:alrN}\,.
\end{equation}
From equations (\ref{eq:alr1}) -- (\ref{eq:alrN}) one infers that only the luminosity-weighted averaged polarization matters. 
Neglecting the uncertainties of energy and luminosity, the error on $\ALR$ is 
\begin{equation}
\Delta \ALR = \sqrt{\frac{1- P^2 \cdot \ALR^2}{N P} + {\ALR^2}\frac{(\Delta P)^2}{P^2}}\,,
\end{equation}
where $P$ is  either $P_{eff}$ if both beams are polarized or  $P_{e^-}$ in case of electron polarization only.
For high statistics $ \Delta \ALR$ is dominated by the uncertainty of the polarization measurement.

\noindent
 Because of error propagation the uncertainty of the effective polarization, $\Delta \peff/\peff$, is considerably smaller than the uncertainty of the individual polarizations $P_{e^+}$ and $P_{e^-}$. In addition, with polarized e$^+$ and e$^-$ the luminosity is effectively increased by a factor $1 + |P_{e^+}P_{e^-}|$. This great advantage can be utilized for measurements if the luminosity-weighted polarizations are equally distributed only to 'opposite' initial states, i.e.\ '$+$ $-$' and '$-$ $+$'. Otherwise, corrections and error propagation would lead to larger errors,  and systematic uncertainties, in particular time-dependent uncertainties, do not cancel. 
The individual polarizations, $P_{e^-}$ and $P_{e^+}$ occur linearly and bi-linearly in equations~(\ref{eq:nonSM}) and (\ref{eq:SM}). That complicates even tiny corrections substantially and requires also the knowledge of correlations between   $P_{e^-}$ and $P_{e^+}$.

\noindent
All these facts emphasize the importance of high  precision polarization  measurement including  high time resolution and cross checks.

\section{Polarized Sources} \label{sec:Pol-source}
\subsection{Polarized Electron Source}\label{sec:ES}
\begin{figure}[h]
\begin{center}
  \includegraphics*[width=0.95\textwidth]{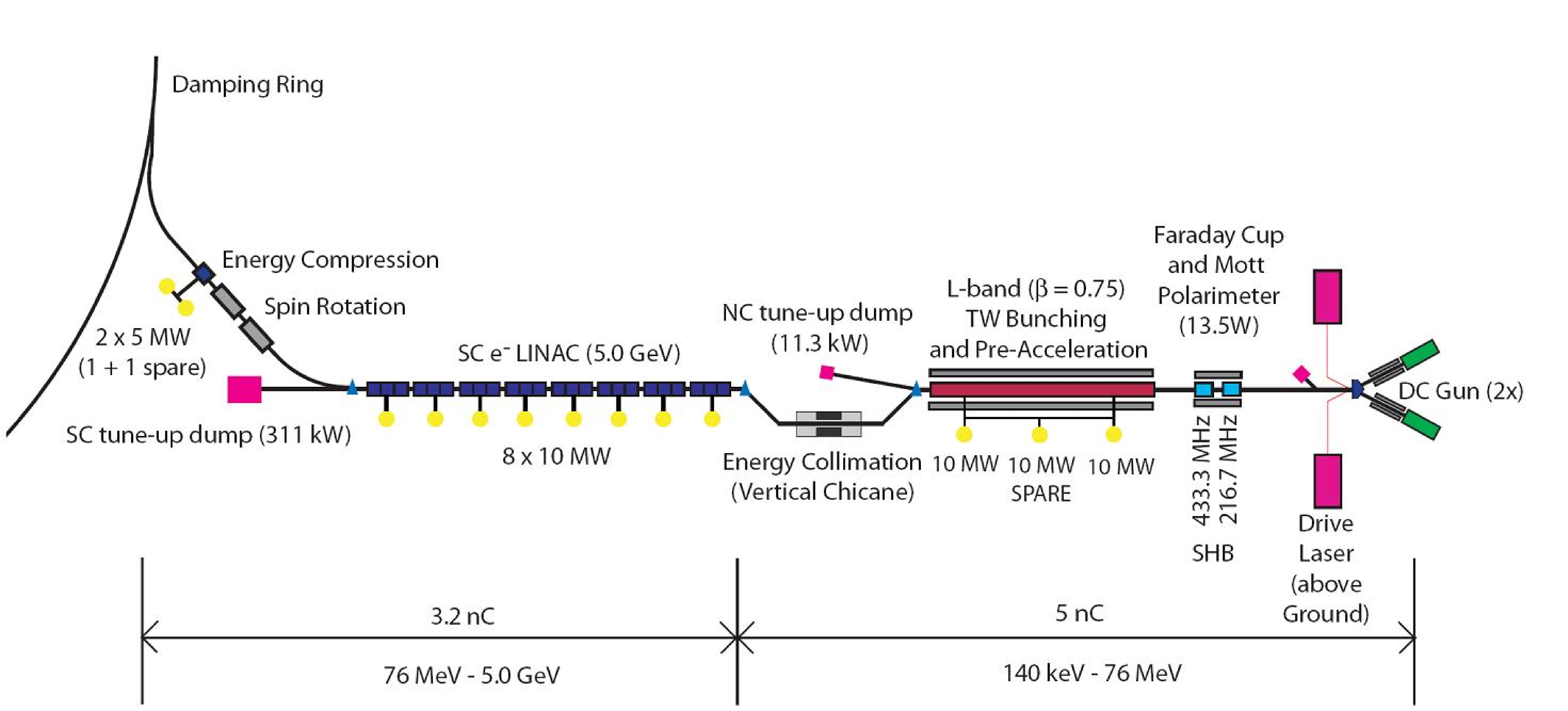}
\caption{Sketch of the electron source as given in the RDR.}\label{fig:ES}
\end{center}
\end{figure}

Polarized electrons are produced from a DC photo-cathode gun. With fast Pockels 
cells the circular polarization of the source laser beam is set and can be reversed train-to-train, thereby allowing  fast reversals of the electron spin. A Mott polarimeter located in a diagnostic line will be used to determine the electron polarization near the source.
Normal-conducting structures are used for bunching and pre-acceleration to 76\,MeV,  afterward the beam is accelerated to 5\,GeV in a superconducting linac. Before injection into the damping ring the spin is rotated to the vertical with superconducting solenoids, the rotation back to the longitudinal is performed before injection to the main linac. A separate superconducting structure is used for bunch compression. The sketch of the electron source as given in the RDR is shown in Figure~\ref{fig:ES}.
Spin rotation at the source could be done also at lower energies as proposed in reference~\cite{ref:400}. The requirements for the solenoid are less stringent for spin rotation at 1.7\,GeV; detailed studies how such system could be implemented have still to be done. Spin rotation  near the gun using a Wien filter is not recommended since substantial emittance growth is expected~\cite{ref:lcws08-brachmann}.

\noindent
Concerning polarization, charge and lifetime, the SLC polarized electron source meets the ILC requirements. But the long bunch trains need a special laser system and normal conducting RF structures that can handle the high RF power. Both requirements are considered as manageable.

\subsection{Polarized Positron Source}\label{sec:PPS}
\begin{figure}[h]
\begin{center}
  \includegraphics*[width=0.95\textwidth]{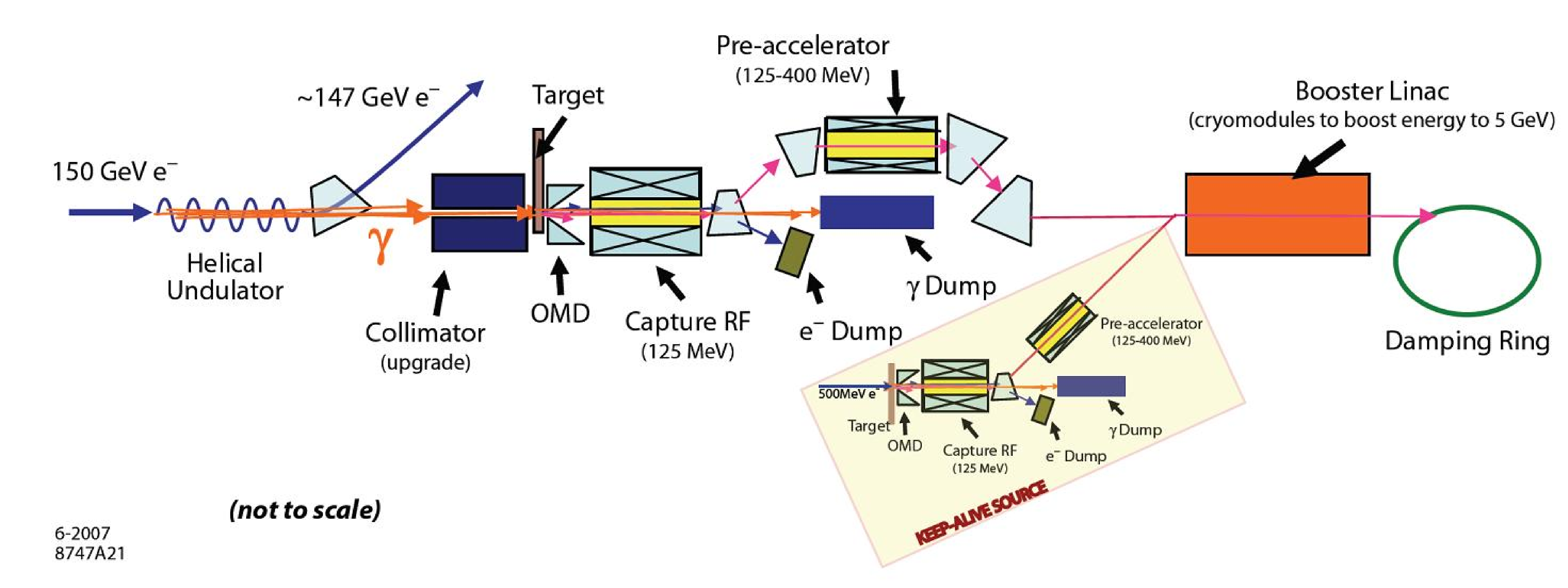}
\caption{Sketch of the positron source as given in the RDR.}\label{fig:PS}
\end{center}
\end{figure}

The positron source uses photo-production to generate 
positrons~\cite{bib:RDRacc}. 
The electron main linac beam passes through a long helical undulator 
to generate a multi-MeV photon beam, which then strikes a thin metal 
target to generate positrons in an electromagnetic shower. 
The positrons are captured, accelerated, separated from the shower 
constituents and unused photon beam and then are transported to the 
Damping Ring. Although the baseline design only requires unpolarized 
positrons, the positron beam produced using a helical undulator 
has a polarization of  $\ppl \apprge 30\,$\%. Simulation studies show that bunch 
(energy) compression would increase the positron capture efficiency 
at the source, with which the positron polarization could even 
reach $\ppl \apprge 45\,$\% at the beginning of the ILC physics 
program.~\cite{bib:e+pol45}. 
Beamline space has been reserved for an upgrade to 60\% polarization.
A sketch of the ILC positron source as described in the RDR is presented in Figure~\ref{fig:PS}.

\noindent
Low energy polarimetry for the positron beam is not foreseen in the RDR.  Different methods and positions for a polarimeter have been studied~\cite{ref:lepol}
based on the special parameters of the positron beam at the source, in particular the relative large transverse beam size. 
Directly after the capture section (125\,MeV) a Compton transmission polarimeter can be used, after pre-acceleration at 400\,MeV a Bhabha polarimeter and after the damping, at 5\,GeV before the main linac, a Compton polarimeter is recommended to measure the positron beam polarization.

\subsubsection{Spin Rotation at the Positron Source}
There are two ways to use the positron beam:
\begin{enumerate}
 \item Physics measurements with a positron polarization 
     of about $\ppl \apprge 30-45\,$\%. 
\item Unpolarized positrons at the $e^+e^-$--IP. 
\end{enumerate}
In the first case (1), the polarized positron beam is transported to the 
$e^+e^-$--IP with minimal spin diffusion and the degree of polarization 
is measured with high precision of 0.25\% near the interaction region 
with upstream and downstream polarimeters (see next sections). 
Spin rotator systems are described in the RDR and are 
included in the positron beam transport lines from the Linac to the Damping 
Ring (LTR) and from the Damping Ring to the Linac (RTL). 
The polarized positrons collide with polarized electrons whose helicity is reversed very frequently (train-by-train). Following the discussion in section~\ref{sec:basics}, it must be possible to flip the positron helicity with the same 
rate. In the current baseline design, however, the positron helicity can 
only be slowly reversed by changing the polarity of the superconducting 
spin rotator magnets. 
As a consequence, the luminosities and the luminosity weighted polarizations are different for the measurements,  systematic uncertainties do not cancel
and finally  the required precision cannot be reached. 
With a positron helicity reversal less frequent than for electrons half of the running time will be spent on the inefficient helicity combinations '$+ ~ + $' and  '$- ~ - $', any gain for the effective luminosity due to positron polarization is lost.

\noindent
Hence it is strongly recommended to modify the baseline configuration to provide random selection of the positron helicity train-by-train by implementing parallel  spin rotator beamlines and kicker systems in the positron   Linac-to-Ring system (LTR) (see also references~\cite{ref:EP08,ref:SR-flip}).
Positron spin rotation and flipping could be done at 5\,GeV~\cite{ref:5000} using superconducting solenoids or at 400\,MeV~\cite{ref:400}.

\noindent
As for the  electron source, at 5\,GeV superconducting solenoids are  necessary to rotate the spin from the transverse horizontal to the vertical direction, at 400\,MeV the solenoid magnets  can be  normal conducting 
and they can be smaller,
demanding less tunnel space. These 
modifications would 
simplify the engineering for these systems, and reduce the costs. But in both cases, at 400\,MeV and at 5\,GeV, two rotation lines are needed for two opposite vertical spin directions. A fast kicker distributes the positron trains to the different lines.

\noindent
In the second case (2),  
the 30-45\% positron polarization are not delivered from the source to the experiment. The spin rotation lines are not needed in the positron line but
a special scheme after the positron damping ring needs to be devised 
to completely destroy the positron polarization in order not to 
adversely effect the physics measurements.
Spin   tracking studies~\cite{ref:larisa} have shown that the  horizontal projections of the spin vectors of an $e^+$ or   $e^-$ bunch do not fully decohere in the damping ring,   even after 8000 turns.
The zero positron polarization also needs to be measured with high 
precision. Further studies are needed to ensure a left-over 
positron DC polarization of about 0.1\% will not affect physics 
measurements, which could result in the need for an even higher 
precision in this case.

\noindent
In both cases, (1) and (2), it is required to measure the positron polarization with high precision, the replacement of the spin rotation and flip facility  
by a device to destroy the polarization to  save costs is not recommended  regarding the physics physics potential with polarized positrons.

\section{Polarimetry near the Interaction Point} \label{sec:Pol-IP}
The ILC offers three methods to measure polarization after acceleration: 
upstream and downstream of the IP, as well as using annihilation events. 
Compton polarimeters are used upstream and downstream.  
The working principle Compton polarimeters can be found for example in reference~\cite{ref:TESLApol}. The longitudinally polarized electron (positron) beam is hit  almost head-on by a circularly polarized laser. The energy spectra of the scattered particles depend on the product of polarizations of laser and lepton beam. The measured asymmetry resulting from polarization reversal is proportional to the beam polarization. 

\noindent
The two polarimeters are highly complementary. The upstream polarimeter 
has a much higher counting rate and time granularity which is important for 
correlation measurements.  The downstream polarimeter has 
access to the depolarization in the interaction. 
 The polarimeters provide corrections and measure the polarization 
on short scales.
Without collisions the two polarimeters can calibrate each other.

\noindent
 With the small errors envisaged at the ILC, it is indispensable for the final polarization measurement at the ILC to have upstream and downstream polarimetry and to get an absolute calibration from annihilation data.  
Cross check of the different ways to measure polarization is mandatory; this has also been confirmed by the polarimetry experience at SLC and by the beam-energy measurements at both, LEP and SLC.

\noindent
To keep the corrections and systematic uncertainties  small, i.e.\ every effort should be made to flip the helicity of electrons and positrons  frequently, if possible train by train.

\subsection{The Upstream Polarimeter}
The upstream polarimeter is located at the beginning of the Beam Delivery System (BDS), upstream 
of the tuneup dump and at a distance of roughly 1.8~km to the $e^+e^-$--IP. 
In this position it benefits from clean beam conditions and very low 
backgrounds compared to any location downstream of the IP. 
It is therefore 
suited to provide very fast and precise polarization measurements 
before collisions. 

\begin{figure}[h]
\begin{center}
  \includegraphics*[width=0.95\textwidth]{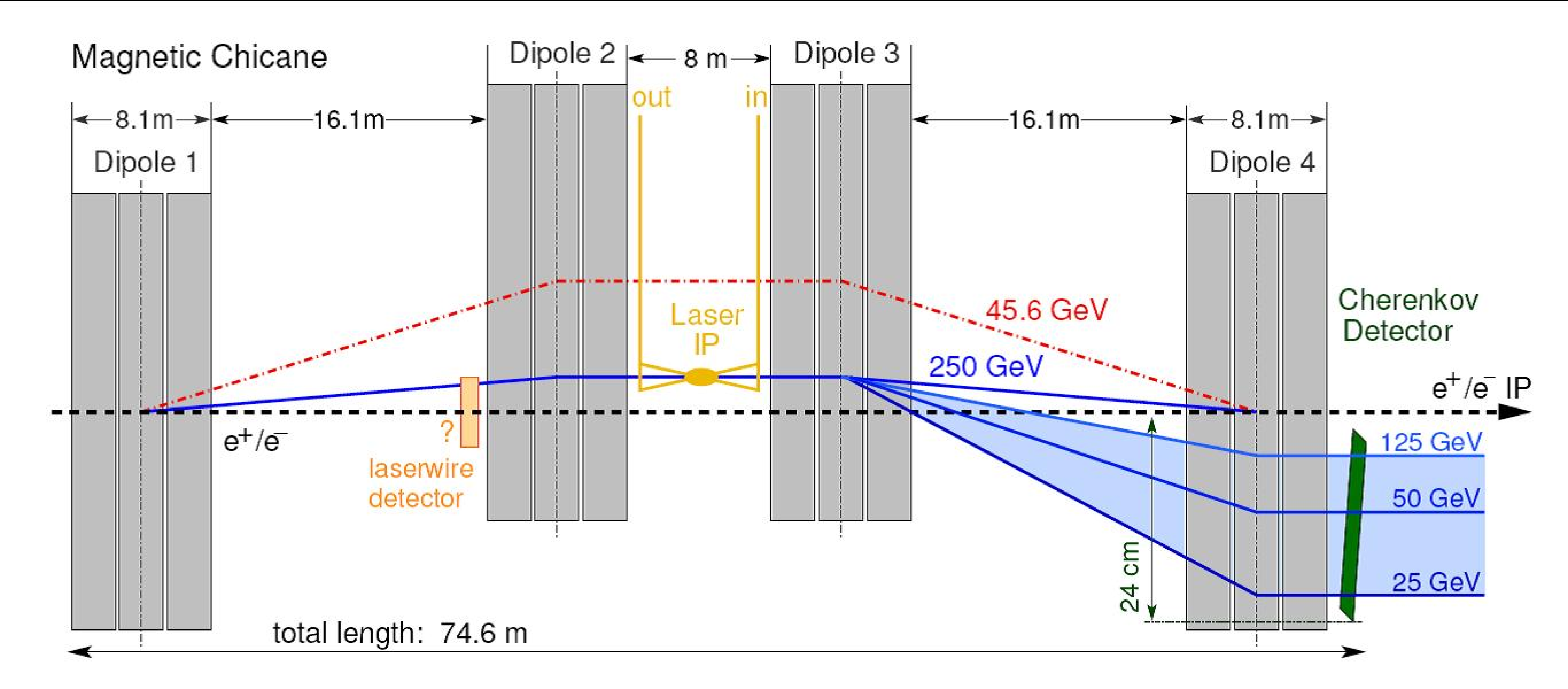}
\caption{Sketch of the upstream polarimeter.}\label{fig:Pol-up}
\end{center}
\end{figure}

\noindent
A complete conceptual layout for the upstream polarimeter had already 
been worked out for TESLA in 2001~\cite{ref:TESLApol}. However, for the ILC, a dedicated chicane-based spectrometer was adopted for upstream polarimetry in 2005,  as this configuration allows the usage of a single laser wavelength at 
all beam energies when the spectrometer is operated with a fixed magnetic 
field. The Compton edge at the detector surface is the same for all beam energies providing a homogeneous detector acceptance and equal performance for all 
center-of-mass energies. 
The layout and principle of the upstream polarimeter chicane is shown in Figure~\ref{fig:Pol-up}: the Compton IP moves laterally with the beam energy; laser, vacuum chamber and detector  have to be designed accordingly.  

\noindent
In this original design with a dedicated fixed-field chicane, the 
upstream polarimeter promised to be a superb and robust instrument with 
broad spectral coverage, very low background, excellent statistical 
performance for all machine bunches, and a high degree of redundancy. 

\noindent
If equipped with a suitable laser, for example a similar one as used
at the FLASH source, 
it can include every single bunch in the measurement. This will permit
virtually instant recognition of variations within each bunch train
as well as time dependent effects that vary train-by-train.
The statistical precision of the polarization measurement will be
already 3\% for any two bunches with opposite helicity, which leads
to an average precision of 1\% for each bunch position in the train 
after the passage of only 20 trains (4 seconds). 
The average over two entire trains with opposite helicity will have 
a statistical error of $\Delta \mathcal{P}/\mathcal{P} = 0.1$\%.

\noindent
The statistical power of the upstream polarimeter depends almost exclusively on the employed laser and therefore to first order factorizes from other design aspects. However the crucial issue which drives the design of the whole polarimeter is to reach an unprecedented low systematic uncertainty of $\delta P /P \leq 0.25\%$ or better~\cite{ref:upstream-prec} with the largest uncertainties coming from the analyzing power calibration (0.2\%) and the detector linearity (0.1\%). 

\noindent
To obtain a useful polarization measurement the beam trajectories are required to be aligned to less than 50~$\mu$rad at the upstream Compton-IP, the collider-IP, and the downstream Compton-IP. This should be achievable by the beam delivery system (BDS) alignment as described in the RDR. 

\noindent
However, the impact of magnets in the interaction region and the crossing angle on the spin alignment needs to be addressed more thoroughly. In the extraction line, corrector magnets are needed to successfully compensate possible deflections resulting from misaligned beam and detector solenoid axes. 

\noindent
In an effort to reduce the cost of the long and expensive BDS system, 
the BDS management decided in autumn 2006 to combine the upstream polarimeter chicane with other diagnostic and machine functions (machine protection system). 
With the machine protection system (MPS) energy collimator in the polarimeter chicane the upstream polarimeter has to be operated  with 'scaled magnetic field' , the excellent polarimeter performance over  the wide ILC energy range is lost.
Another suggestion is laser wire emittance diagnostics with a detector in front of the second dipole triplet of the polarimeter chicane as can be seen also in Figure~\ref{fig:Pol-up}. It is expected that this detector creates huge background unacceptable for polarimetry.
 The consequences of combining the upstream polarimeter chicane with other diagnostic and machine functions were discussed at the Workshop on Polarization and Energy Measurement in April in Zeuthen~\cite{ref:EP08} and  at this  workshop~\cite{ref:jenny-pol-lcws08}.  
Better solutions will be found.

\subsection{The Downstream Polarimeter}
The downstream polarimeter is located about 150~m downstream of the $e^+e^-$--IP in the extraction line and on axis with the IP and IR magnets. It can measure the beam polarization both with and without collisions, thereby testing the calculated depolarization correction which is expected to be at the 0.1-0.2\% level. 
A complete conceptual layout for the downstream polarimeter exists, including 
magnets, laser system and detector configuration. 
\begin{figure}[h]
\begin{center}
  \includegraphics*[width=0.95\textwidth]{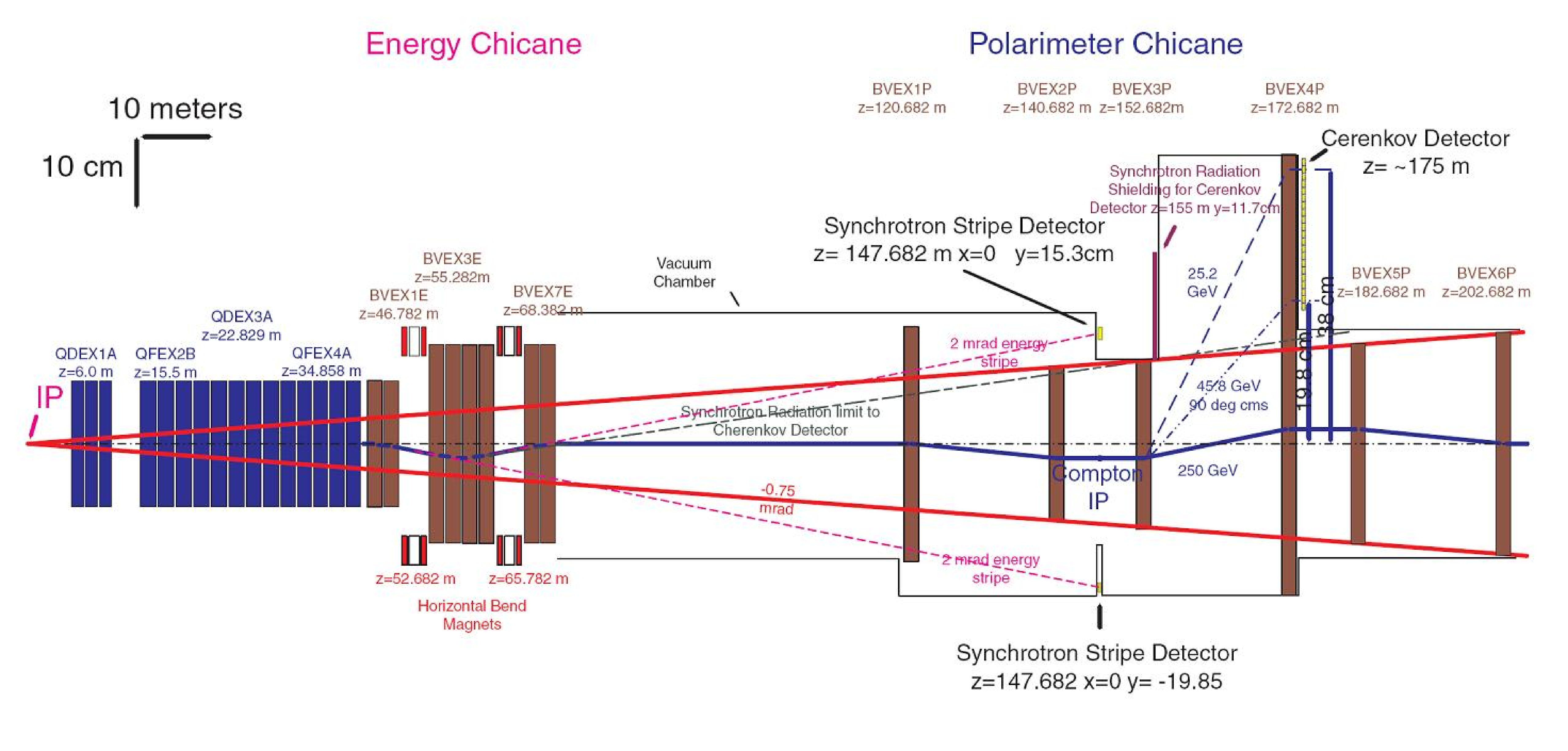}
\caption{Sketch of the downstream polarimeter.}\label{fig:Pol-down}
\end{center}
\end{figure}

\noindent
Figure~\ref{fig:Pol-down} presents the sketch of the extraction line polarimeter as shown in the RDR. The  downstream polarimeter needs a six-magnet chicane~\cite{bib:Moffeit-6magnets}; the additional two  magnets after the Compton detector allow to operate the third and fourth magnets at higher field to deflect the Compton electrons further from the beam line and to return the beam to the nominal trajectory. 

\noindent
Three 10~Hz laser systems can achieve Compton collisions for three out of 2800 bunches in a train. 
Each laser will sample one particular bunch in a train for a time interval of 
a few seconds to a minute, then select a new bunch for the next time interval, and so on in a pre-determined pattern. 
The Compton statistics are high with about 300 Compton scattered electrons 
per bunch in a detector channel at the Compton edge. 

\noindent
With this design, a statistical uncertainty of less than 1\% per minute can be achieved for each of the measured bunches. This is dominated by fluctuations in Compton luminosity due to beam jitter and laser targeting jitter and to possible background fluctuations. The statistical error due to Compton statistics in one minute, for a bunch sampled by one laser, is 0.3\%.
However, if compared to the average precision of the upstream polarimeter, 
a similar precision for each bunch position in a train could only be reached 
after about 17 hours.

\noindent
Background studies have been carried out for disrupted beam losses and for the influence of synchrotron radiation. There are no significant beam losses for the nominal ILC parameter set and beam losses look acceptable even for the low power option. A synchrotron radiation collimator protects the Compton detector and no significant synchrotron radiation backgrounds are expected.
The systematic precision is expected to be about 0.25\%, with the largest 
uncertainties coming from the analyzing power calibration (0.2\%) and 
detector linearity (0.1\%).

\noindent
In addition, it is desired to correlate the downstream polarimeter measurements with  beam position measurements (BPM)  at the $e^+e^-$--IP~\cite{bib:EPWS-Woods-MDIissues} by providing information from the  BPM system to the 
polarimeter DAQ including bunch number identification.

\subsection{Depolarization at the IP}\label{sec:depol}
For high precision measurements depolarization effects in the machine have to be known. Depolarization occurs during bunch crossing because of beam-beam effects or could be caused by misalignment, i.e.\ ground motion induced misalignment.
Depolarization effects and their impact on physics have to be understood with high precision, for details see reference~\cite{ref:hartin}.

\noindent
With the downstream polarimeter depolarization effects can be measured by comparing measurements with colliding and non-colliding beams and simulations.
The disrupted beam is propagated down the the extraction line to simulate the the polarization measurement with the downstream polarimeter.     

\noindent
Current studies using Guinea Pig++~\cite{ref:GP} and CAIN~\cite{ref:cain} show good agreement. 
 For a nominal case in which both beams are initially assumed to be 100\%
longitudinally polarized, the luminosity-weighted polarization of each
beam is predicted to be reduced by $0.21\%\pm0.01\%$ compared to the
initial polarization. For details see references~\cite{ref:rimbault,ref:bailey}.

\subsection{Polarimetry with Annihilation Data}\label{sec:annih}
Apart from the polarimeters, polarization can also be measured using annihilation  data~\cite{bib:PolScheme-Moenig}. 
If both beams are polarized this can be done measuring the fermion-pair cross section for different helicity combinations in the initial state (Blondel scheme). If only the electron beam is polarized,  W-pair production can be used to determine electron polarization with the assumption that the $e\nu W$-coupling is purely left-handed which is well tested. 

\subsubsection{Blondel Scheme}
The annihilation  method using fermion pairs is based on equations (\ref{eq:alr1}). By measuring all four cross sections, $\sigma_{\pm \mp}$ and $\sigma_{\pm \pm}$, a direct and independent measurements of the luminosity weighted  electron and positron polarization as well as the asymmetry $\ALR$ are possible. Nevertheless, the polarimeters near the IP are needed to monitor time dependent  fluctuations in the polarizations. 
This
Blondel-scheme is especially efficient at the Z resonance where a high rate of annihilation data is available and the polarization has to be measured  with extremely high precision. 
At the Z peak, only a small fraction of the luminosity ($\sim 10$\%) has to be delivered to the 'inefficient' helicity combinations '$\pm~ \pm$' but half of the  luminosity at high energies~\cite{bib:PolScheme-Moenig}.

Only a small fraction of the luminosity ($\sim 10$\%) has to be delivered to the 'inefficient' helicity combinations '$\pm~ \pm$'. At high energies half of the  luminosity should be deliverd to the  helicity combinations '$\pm~ \pm$' to reach the required  uncertainties for the polarization measurement. Due to the low event rates the polarization measurement with the Blondel Scheme is slow but it is a good cross check for the  measurement with polarimeters. A detailed discussion of the Blondel scheme can be found in reference~\cite{ref:blondel}.

\subsubsection{{\boldmath$Z$}-pole Calibration Data}

\noindent
The GigaZ option -- ILC  precision measurements at the $Z$-pole -- is foreseen as a running mode at a later time of ILC operation. However, calibration will require runs at the Z peak. Measurements during these runs are far beyond the statistics of GigaZ but will exceed that of LEP and provide complementary information to the presently known precision measurement of the weak mixing angle $\sin \theta_{eff}$. 

\noindent  
Hence the  precise measurements of energy and polarization at the $Z$-pole are important  and should be included in the ILC baseline documents for the following reasons: 
\begin{itemize}
\item  Polarimeter calibration and cross-check against physics based 
  polarization measurements using the Blondel scheme;
\item  Data from these calibration runs can also provide 
  significant statistics for physics measurements complementary to LEP and SLC results.
\end{itemize}
Reference~\cite{bib:Zpole-Gudi} summarizes the physics prospects with $Z$-pole calibration data.

\subsubsection{Polarization Measurement with W$^+$W$^-$ Data}
Also the W$^+$W$^-$ data can be used for polarization measurement~\cite{bib:PolScheme-Moenig}.
The forward peak is entirely dominated by t-channel neutrino exchange and not 
influenced by possibly unknown triple gauge interactions.
Providing long measuring times (months!) a  precision of  0.1\% can be reached. Nevertheless, the polarimeters are needed to  correct for time-dependent relative polarization fluctuations.

\section{Upgrade to {\boldmath $\sqrt{s} = 1$}\,TeV}
Depending on the  LHC physics results an energy upgrade to 1\,TeV center-of-mass will be the next step after the completion of the  baseline programme. The requirements for polarimetry are the same as for $E_{cms}=500\,$GeV. 
The upstream polarimeter with the fixed field operation allows for precise  polarization measurements at 1\,TeV. This performance should  not be compromised in any way. In particular, the relatively easy upgrade to operate polarimeters at  beam energies above 250\,GeV should not be rendered impossible by installed  beam diagnostic or other instrumentation in the chicane.

\section{Recommendations form the EP Workshop in Zeuthen, April 2008}\label{sec:recos}
At the ``Workshop on Energy and Polarization Measurement'' in April 2009 at DESY Zeuthen seven recommendations emerged requiring follow-up from the GDE and the Research  Director (see also reference~\cite{ref:EP08}). All these items were also discussed at LCWS08/ILC08.
  \begin{enumerate}
  \item  Separate the functions of the upstream polarimeter chicane. 
    Do not include an MPS energy collimator or laser-wire emittance 
    diagnostics; use instead a separate 
    setup for these two. 
  \item  Modify the extraction line polarimeter chicane from a 4-magnet 
    chicane to a 6-magnet chicane to allow the Compton electrons to be 
    deflected further from the disrupted beam line.
  \item  Include precise polarization and beam energy measurements 
    for {\boldmath$Z$}-pole calibration runs into the baseline configuration.
  \item  Keep the initial positron polarization of 30-45\% for physics (baseline).
  \item  Implement parallel spin rotator beamlines with a kicker system before 
    the damping ring to provide rapid helicity flipping of the positron spin. 
  \item  Move the pre-DR positron spin rotator system from 5~GeV to 400~MeV. 
    This eliminates expensive superconducting magnets and reduces costs.
  \item  Move the pre-DR electron spin rotator system to the source area. 
    This eliminates expensive superconducting magnets and reduces costs.
  \end{enumerate}

\section{Summary}
The studies, talks and discussions presented  at this conference 
demonstrated that beam polarization and its measurement are crucial for the physics success of any future linear collider.
To achieve the  required precision it is absolutely decisive to employ 
multiple devices for testing and controlling the systematic uncertainties of 
each polarimeter. 
The polarimetry methods for the ILC are complementary: with the upstream polarimeter the measurements are performed in a clean environment, they are fast and allow to monitor time-dependent variations of polarization. The polarimeter downstream the IP will measure the disrupted beam resulting in high background and much lower statistics, but  it allows access to the depolarization at the IP.
Cross checks between the polarimeter results give redundancy and inter-calibration  which is essential for high precision measurements. 
Current plans and issues for polarimeters and also energy spectrometers in the Beam Delivery System of the ILC are summarized in reference~\cite{ref:LoI}.

\noindent
The ILC baseline design allows already from the beginning the operation with polarized electrons and polarized positrons provided the spin rotation and the fast helicity reversal for positrons will be implemented. A reversal of the positron helicity significantly slower than that of electrons is not recommended to not   compromise the precision and hence the success of the ILC. 

\noindent
Recently to use calibration data at the Z resonance for physics has been discussed. It looks promising but further  studies are needed to evaluate and to optimize these measurements.

\noindent
Finally it should be remarked: many studies on different physics processes and scenarios at a future linear collider are done for high luminosities and high energy assuming small and well-known uncertainties. Polarization, especially positron polarization, is often considered as not that important. But in order to interpret data and to reduce ambiguities in the measurements, the polarization of electrons and positrons and their very precise knowledge are essential. The ILC design must  offer this from the beginning  to be prepared for the physics questions  after years of LHC operation.


\begin{footnotesize}

\end{footnotesize}

 

\begin{thebibliography}{99}
\bibitem{url}
S. Riemann, Presentation at LCWS08 \\ {\scriptsize
\verb$http://ilcagenda.linearcollider.org/contributionDisplay.py?contribId=54&sessionId=2&confId=2628$}
 
%
\bibitem{ref:power}
  G.A.~Moortgat-Pick {\it et al.},
  ``Polarised Positrons and Electrons at the Linear Collider'', 
  Phys.\ Rept.\ {460} (2008) 131; 
  [arXiv:hep-ph/0507011].

%
\bibitem{bib:RDRacc}
  ILC Global Design Effort and World Wide Study, 
  Editors: N.~Phinney, N.~Toge, and N.~Walker, 
  ``International Linear Collider Reference Design Report - Volume 3: Accelerator'',
  August 2007.

%
\bibitem{ref:spin-rot}
  P.~Schmid,
  ``A spin rotator for the ILC'', EUROTEV-REPORT-2005-024;\\
  P.~Schmid and N.~J.~Walker, ``A spin rotator for the ILC'',
  EUROTEV-REPORT-2006-068.

%
\bibitem{ref:jenny-pol-lcws08}
  J.~List and D.~K\"afer,
  ``Improvements to the ILC Upstream Polarimeter'',
  arXiv:0902.1516 [physics.ins-det].

%
\bibitem{ref:EP08}
  B.~Aurand {\it et al.},
  ``Executive Summary of the Workshop on Polarization and Beam Energy
  Measurements at the ILC'',
  arXiv:0808.1638 [physics.acc-ph].

%
\bibitem{ref:RDRphys}
  ILC Global Design Effort and World Wide Study, 
  Editors: A.~Djouadi, J.~Lykken, K.~M\"onig, Y.~Okada, M.~Oreglia, and S.~Yamashita, 
  ``International Linear Collider Reference Design Report - Volume 2: Physics at the ILC'', 
  August 2007.

%
\bibitem{ref:pol08} 
  C.~Bartels, C.~Helebrant, D.~K\"afer and J.~List,
  ``Compton Cherenkov Detector Development for ILC Polarimetry'',
  arXiv:0902.3221 [physics.ins-det].

%
\bibitem{ref:400}
K.~ Moffeit, M.~Woods, D.~Walz,
``Spin Rotation at lower energy than the damping ring'', 
ILC-NOTE-2008-040.

%
\bibitem{ref:lcws08-brachmann}
A.~Brachmann, Presentation at ILC08 \\ {\scriptsize
\verb$http://ilcagenda.linearcollider.org/contributionDisplay.py?contribId=250&sessionId=10&confId=2628$}

%
\bibitem{bib:e+pol45}
  A.~Ushakov, 
  Talk given at the Positron Source Coll. Meeting, 
  April 7-9, 2008 at DESY in Zeuthen; \\
  \verb$http://ilcagenda.linearcollider.org/contributionDisplay.py?contribId=13&confId=2639$;
  Y.K.~Batygin, 
  ``Spin Rotation and Energy Compression in the ILC Linac-to-Ring Positron Beamline'', 
  Nucl.\ Instrum.\ Meth.\ A 570 (2007) 365.

%
\bibitem{ref:lepol}
  G.~Alexander{\it  et al},
  ``Low-energy Positron Polarimetry at the ILC''
  EUROTeV-Report-2008-091.

%
\bibitem{ref:SR-flip} 
  S.~Riemann, A.~Sch\"alicke and A.~Ushakov,
  ``Frequency of Positron Helicity Reversal'',
  arXiv:0903.2366 [physics.ins-det].

%
\bibitem{ref:5000}
  K.~Moffeit, M.~Woods, P.~Sch\"uler, K.~M\"onig and P.~Bambade,
  ``Spin rotation schemes at the ILC for two interaction regions and  positron
  polarization with both helicities'', SLAC-TN-05-045.

%
\bibitem{ref:larisa}
  L.~Malysheva,
 ``Depolarisation in the damping rings of the ILC'',
{\it In the Proceedings of 2007 International Linear Collider Workshop (LCWS07 and ILC07), Hamburg, Germany, 30 May - 3 Jun 2007, pp DR003}.

%
\bibitem{ref:TESLApol}
  V.~Gharibyan, N.~Meyners and P.~Sch\"uler,
  ``The TESLA Compton polarimeter'',
  LC-DET-2001-047.

%
\bibitem{ref:upstream-prec}
  C.~Helebrant, D.~K\"afer and J.~List,
  ``Precision Polarimetry at the International Linear Collider'',
  arXiv:0809.4485 [physics.ins-det].

%
\bibitem{bib:Moffeit-6magnets} 
  K.~Moffeit {\it et al.}, 
  ``Proposal to modify the polarimeter chicane in the ILC 14 mrad extraction line'', 
  SLAC-PUB-12425, IPBI\_TN-2007-1, March 2007.  

%
\bibitem{bib:EPWS-Woods-MDIissues}
  M.~Woods,
  ``Machine-Detector Interface Issues for the ILC Polarimeters'',
  SLAC-PUB-13259.

%
\bibitem{ref:hartin}
A.~Hartin, ``Depolarization from the upstream to the downstream polarimeter'' \\ {\scriptsize
\verb$http://ilcagenda.linearcollider.org/contributionDisplay.py?contribId=307&sessionId=13&confId=2628$}, to appear in these proceedings.

%
\bibitem{ref:GP} 
D.~Schulte, Ph. D. Thesis, University of Hamburg 1996.
TESLA-97-08,
K. Yokoya, P. Chen, ``Beam-beam phenomena in linear colliders'',
KEK Preprint 91-2, April 1991.

%
\bibitem{ref:cain}
K.~Yokoya, P.~Chen, SLAC-PUB-4692, 1988; K.~Yokoya, ``User's Manual of CAIN'', Version 2.35, April 2003.

%
\bibitem{ref:rimbault}
C.~Rimbault, ``Implementation and study of depolarising effects in the GP++ beam$\rightarrow$ beam interaction simulation'' \\ {\scriptsize
\verb$http://ilcagenda.linearcollider.org/contributionDisplay.py?contribId=308&sessionId=24&confId=2628$},\\
to appear in these proceedings.

%
\bibitem{ref:bailey}
I.R.~Bailey {\it et al.},  
  ``Depolarization and Beam-Beam Effects at the Linear Collider'', 
  EUROTEV-REPORT-2008-026.

%
\bibitem{bib:PolScheme-Moenig} 
  K.~M\"onig,
  ``Polarisation Measurements with Annihilation Data'',
  Proceedings of LCWS,
  Paris, April 2004, Vol.~2, 875, \'Edition de l'\'Ecole Polythechnique.

%
\bibitem{ref:blondel}
  K.~M\"onig, ``Electroweak physics at a linear collider Z-factory'',
  LC-PHSM-1999-002-TESLA;\\
  K.~M\"onig,
  ``The use of positron polarization for precision measurements'',
  LC-PHSM-2000-059.

%
\bibitem{bib:Zpole-Gudi} 
  G.~Moortgat-Pick {\it et al.}, 
  ``Precision Measurements with Calibration Data at the $Z$-pole'',
  Paper in preparation.   

\bibitem{ref:LoI}
  S.~Boogart {\it et al.},
  ``Polarimeters and Energy Spectrometers for the ILC Beam Delivery System'',
  ILC-Note-2009-049.
\end{thebibliography}
\end{document}